\documentstyle[psfig,12pt]{article}

\def\thefootnote{\fnsymbol{footnote}}
\def\Tr{{\rm Tr}}
\def\ignorethis#1{}
\def\[{\left [}
\def\]{\right ]}
\def\({\left (}
\def\){\right )}
\def\lbr{\left\{}
\def\rbr{\right\}}
\newcommand{\dilaton}{{\ell}}
\newcommand{\Lag}{{\cal L}}
\newcommand{\superint}{\int \diff^{4}\theta}
\newcommand{\lowest}{|_{\theta =\bar{\theta}=0}}
\newcommand{\diff}{\mbox{d}}

\newcommand{\WaWa}{\Tr({\cal W}^{\alpha}{\cal W}_{\alpha})}

\newcommand{\DaDa}{{\cal D}^{\alpha}{\cal D}_{\alpha}}
\newcommand{\DbDb}{{\cal D}_{\dot{\alpha}}{\cal D}^{\dot{\alpha}}}

\newcommand{\ReT}{{\(T^{I}+{\overline{T}}^{I}\)}}
\newcommand{\baal}{b_{a}^{\alpha}}
\newcommand{\baaleff}{\(\baal\)_{\rm eff}}
\newcommand{\bpaleff}{\(b_{+}^{\alpha}\)_{\rm eff}}
\newcommand{\caleff}{\({c_{\alpha}}\)_{\rm eff}}
\newcommand{\lang}{\left\langle}
\newcommand{\rang}{\right\rangle}
\newcommand{\order}{{\cal O}}

\begin{document}
\begin{titlepage}
\begin{center}
            \hfill    LBNL-44305 \\
            \hfill    UCB-PTH-99/44 \\
            \hfill    hep-ph/yymmxxx \\[0.03in]
\vskip .2in
{\large \bf Constraints on Hidden Sector Gaugino Condensation}
\footnote{This work was supported in part by the Director, Office of 
Energy Research, Office of High Energy and Nuclear Physics, Division 
of High Energy Physics of the U.S. Department of Energy under 
Contract DE-AC03-76SF00098  and in part by the National Science 
Foundation under grant PHY-95-14797 and PHY-94-04057.}\\[.1in]

Mary K. Gaillard 
and
Brent D. Nelson \\[.05in]

{\em  Theoretical Physics Group \\
      Ernest Orlando Lawrence Berkeley National Laboratory \\
      University of California, Berkeley, California 94720 \\
      and \\
      Department of Physics \\
      University of California, Berkeley, California 94720}\\[.1in]
\end{center}

\begin{abstract}
We study the phenomenology of a class of models describing modular
invariant gaugino condensation in the hidden sector of a low-energy
effective theory derived from the heterotic string. Placing simple
demands on the resulting observable sector, such as a
supersymmetry-breaking scale of approximately 1 TeV, a vacuum with properly
broken electroweak symmetry, superpartner masses above current direct
search limits, etc., results in significant restrictions on the
possible configurations of the hidden sector.

\end{abstract}
\end{titlepage}
\renewcommand{\thepage}{\roman{page}}
\setcounter{page}{2}
\mbox{ }

\vskip 1in

\begin{center}
{\bf Disclaimer}
\end{center}

\vskip .2in

\begin{scriptsize}
\begin{quotation}
This document was prepared as an account of work sponsored by the United
States Government.  Neither the United States Government nor any agency
thereof, nor The Regents of the University of California, nor any of their
employees, makes any warranty, express or implied, or assumes any legal
liability or responsibility for the accuracy, completeness, or usefulness
of any information, apparatus, product, or process disclosed, or represents
that its use would not infringe privately owned rights.  Reference herein
to any specific commercial products process, or service by its trade name,
trademark, manufacturer, or otherwise, does not necessarily constitute or
imply its endorsement, recommendation, or favoring by the United States
Government or any agency thereof, or The Regents of the University of
California.  The views and opinions of authors expressed herein do not
necessarily state or reflect those of the United States Government or any
agency thereof of The Regents of the University of California and shall
not be used for advertising or product endorsement purposes.
\end{quotation}
\end{scriptsize}

\vskip 2in

\begin{center}
\begin{small}
{\it Lawrence Berkeley Laboratory is an equal opportunity employer.}
\end{small}
\end{center}

\newpage
\renewcommand{\thepage}{\arabic{page}}
\def\thefootnote{\arabic{footnote}}
\setcounter{page}{1}
\setcounter{footnote}{0}
%THIS IS PAGE 1 (INSERT TEXT OF REPORT HERE)

When considering the subject of effective field theories from strings, 
the notion of ``phenomenological viability'' has in the past been a
very loose standard. Indeed some of the well-known problems facing
such low-energy theories seemed quite intractable, depressing the
prospects of ever being able to refer to a meaningful superstring
phenomenology. The problems to which we refer include the need to
generate a hierarchy between the supersymmetry-breaking scale and the
Planck scale, the cosmological dangers of moduli fields with
Planck-suppressed interactions, the desire for a
weakly-coupled effective quantum field theory, and most significantly the
need to stabilize the dilaton \cite{dilaton}.

Recently \cite{DilStab,ModInv,susybreak}, however, it was shown
that by incorporating postulated
nonperturbative string-theoretical effects in a modular invariant
low-energy field theory the above problems can be addressed in a
simple manner with tuning required only in the vanishing of the
cosmological constant. Having passed these initial tests it now
becomes possible to ask for a slightly higher standard in
``viability.'' 

The philosophy behind this study is to probe this class
of models in a series of phenomenological arenas to uncover relations
between the dynamics of the hidden sector and the
nature of our observable world. After a
review in Section~\ref{sec:model} of the class
of models previously developed in
\cite{DilStab,ModInv,susybreak,gauginomass}, we investigate in
Section~\ref{sec:phenom} the
initial challenge of setting the supersymmetry-breaking scale that all
effective field theories from strings must
confront. This is largely a reiteration of results discussed in
\cite{susybreak}. In Section~\ref{sec:RGE} we turn to the pattern of
soft supersymmetry-breaking parameters and look for the implications of current 
mass bounds arising from searches at LEP and the Tevatron. Finally,
Section~\ref{sec:gaugeunify} considers the question of gauge coupling
unification in the context of string theory.
\pagebreak
\section{Model}
\label{sec:model}
\subsection{The Effective Lagrangian}
\label{sec:Lag}
The following is a condensation of material more fully presented in
\cite{ModInv,gauginomass} and aims to bring together the key points necessary for
the subsequent discussion of phenomenological consequences. In those
references, as here, the K\"ahler $U(1)$
superspace formalism of \cite{BGG} is used throughout.

Supersymmetry breaking is implemented via condensation of gauginos
charged under the hidden sector gauge group ${\cal G}=\prod_{a}{\cal
  G}_{a}$, which is taken to be a subgroup of $E_8$. For each gaugino
condensate a vector superfield $V_a$ is introduced and the gaugino
condensate superfields $U_{a}\simeq {\WaWa}_{a}$ are then identified as
the (anti-)chiral projections of the vector superfields: 
\begin{eqnarray}
U_{a}=-\(\DbDb-8R\)V_{a},& {\overline{U}}_{a}=-\(\DaDa-8\overline{R}\)V_{a}.
\label{eq:chiralproj}
\end{eqnarray}
The dilaton field (in the linear multiplet formalism \cite{linear}
used here) is the lowest component of the vector superfield
$V={\sum_{a}}V_{a}$: $\dilaton =V{\lowest}$. Note that none of the individual
lowest components ${V_a}\lowest$ will appear in the effective theory
component Lagrangian.

In the class of orbifold compactifications we will be considering
there are three untwisted 
moduli chiral superfields $T^{I}$ and matter chiral superfields
$\Phi^{A}$ with K\"ahler potential
\begin{eqnarray}
K=k\(V\)+\sum_{I}g^{I} +{\sum_A}e^{{\sum_I}{q_{I}^{A}}{g^{I}}} \left| \Phi^{A}
\right|^{2} + \order \( \Phi^{4} \), & g^{I}=-\ln{\ReT},
\label{eq:Kahlerpot}
\end{eqnarray}
where the $q_{I}^{A}$ are the modular weights of the fields
$\Phi^{A}$. The relevant part of the complete effective Lagrangian is then
\begin{equation}
{\Lag}_{{\rm eff}}={\Lag}_{{\rm KE}} + {\Lag}_{{\rm VY}} +
{\Lag}_{{\rm pot}} + {\sum_{a}}{\Lag}_{{\rm a}} + {\Lag}_{{\rm GS}},
\label{eq:completeLag}
\end{equation}
where
\begin{eqnarray}
{\Lag}_{{\rm KE}}={\superint}E\[-2+f\(V\)\], & k\(V\)=\ln{V}+g\(V\),
\label{eq:LKE}
\end{eqnarray}
is the Lagrangian density for the gravitational sector coupled to the
vector multiplet and gives the kinetic energy terms for the dilaton,
chiral multiplets, gravity superfields and tree-level Yang-Mills terms.
Here the functions $f\(V\)$ and $g\(V\)$ represent nonperturbative
corrections to the K\"ahler potential arising from string effects. The 
two functions $f$ and $g$ are related by the requirement that the
Einstein term in~(\ref{eq:LKE}) have canonical normalization:
\begin{equation}
V\frac{\diff g\(V\)}{\diff V}=-V\frac{\diff f\(V\)}{\diff V}+f\(V\),
\label{eq:fgrel}
\end{equation}
and obey the weak-coupling boundary conditions:
$f\(0\)=g\(0\)=0$. In the {\mbox presence} of these nonperturbative effects
the relationship between the dilaton and the effective field
theory gauge coupling becomes 
${g^2}/{2}={\dilaton}/{\(1+f\(\dilaton\)\)}$.

The second term in~(\ref{eq:completeLag}) is a generalization of the original
Veneziano-Yankielowicz superpotential term \cite{VY}, 
\begin{equation}
{\Lag}_{\rm VY}=\frac{1}{8}{\sum_a}{\superint}\frac{E}{R}{U_a}
\left [ {b_{a}'} \ln{\(e^{-K/2}{U_a}\)} + {\sum_{\alpha}} {b_{a}^{\alpha}} 
\ln{\[ \( {\Pi^{\alpha}} \)^{p_{\alpha}} \] } \right ] + {\rm h.c.},
\label{eq:LagVY}
\end{equation}
which involves the gauge
condensates $U_a$ as well as possible gauge-invariant matter condensates described by
chiral superfields $\Pi^{\alpha} \simeq {\prod_{A}} {\( \Phi^{A}
  \)}^{n_{\alpha}^{A}}$~\cite{VYmod}. Neither the gaugino nor the matter
condensate superfields are taken to be propagating~\cite{yiyen}. The
coeffecients $b_{a}'$, $\baal$ and $p_{\alpha}$ are determined
by demanding the correct transformation properties of the expression
in~(\ref{eq:LagVY}) under chiral and conformal
transformations~\cite{ModInv,match} and yield the following relations:
\begin{eqnarray}
b_{a}'=\frac{1}{8{\pi}^2}\( C_a - \sum_{A}C_{a}^{A} \), &
{\displaystyle {\sum_{\alpha,A}} {b_{a}^{\alpha}} {n_{\alpha}^{A}}{p_{\alpha}} =
{\sum_{A}}\frac{C_{a}^{A}}{4{\pi}^2}}, & {p_{\alpha}}\sum_{A}
n_{\alpha}^{A}=3 \ \ \ \forall \alpha.
\label{eq:coeff}
\end{eqnarray}
The final condition amounts to choosing the value of $p_{\alpha}$ so
that the effective operator $\( \Pi^{\alpha} \) ^{p_{\alpha}}$ has
mass dimension three. In~(\ref{eq:coeff}) the quantities $C_{a}$ and
$C_{a}^{A}$ are the quadratic Casimir operators for the adjoint and matter representations,
respectively. Given the above relations it is also convenient to
define the combination
\begin{equation}
b_{a}\equiv b_{a}' + {\sum_{\alpha}}\baal =\frac{1}{8{\pi}^2}
\(C_{a}-\frac{1}{3} {\sum_A}C_{a}^{A}\)
\label{eq:betafunc}
\end{equation}
which is proportional to the one-loop beta-function coefficient for
the condensing gauge group ${\cal G}_a$.

The third term in~(\ref{eq:completeLag}) is a superpotential term for
the matter condensates consistent with the symmetries of the
underlying theory
\begin{equation}
{\Lag}_{\rm pot}=\frac{1}{2}{\superint}\frac{E}{R}{e^{K/2}}W\[
\( \Pi^{\alpha}\)^{p_{\alpha}} ,T^{I} \] + {\rm h.c.}.
\label{eq:Lagpot}
\end{equation}
We will adopt the same set of simplifying assumptions taken up
in~\cite{ModInv}, namely that for fixed $\alpha$, $\baal\neq 0$ for only
one value of $a$. Then $u_{a}=0$ unless $W_{\alpha} \neq 0$ for every
value of $\alpha$ for which $\baal\neq 0$. We next assume that there
are no unconfined matter fields charged under the hidden sector gauge
group and ignore possible dimension-two matter condensates involving
vector-like pairs of matter fields. This allows a simple factorization 
of the superpotential of the form
\begin{equation}
W\[ \({\Pi}^{p_{\alpha}}\),T \]={\sum_{\alpha}}{W_{\alpha}}\(T\)\(
{\Pi}^{\alpha} \)^{p_{\alpha}},
\end{equation}
where the functions $W_{\alpha}$ are given by
\begin{equation}
{W_{\alpha}}\(T\)=c_{\alpha}{\prod_I}\[{\eta}\(T^{I}\)\]^{2\({p_{\alpha}} 
  {q^{\alpha}_{I}}-1\)}.
\label{eq:LagPot}
\end{equation}
Here ${q^{\alpha}_{I}}={\sum_{A}}{n_{\alpha}^{A}}{q_{I}^{A}}$ and 
the Yukawa coefficients $c_{\alpha}$, while {\it a priori} unknown variables, are 
taken to be of $\order\(1\)$. The function $\eta ( T^{I})$ is the
Dedekind function and its presence in~(\ref{eq:LagPot}) ensures the
modular invariance of this term in the Lagrangian.

The remaining terms in~(\ref{eq:completeLag}) include the quantum
corrections from light field loops to the unconfined
Yang-Mills couplings and the
Green-Schwarz (GS) counterterm introduced to ensure modular
invariance.\footnote{Not included in this paper are
  string loop corrections $\( {\Lag}_{\rm th}\)$ \cite{N2loop} which
  vanish for orbifold compactifications 
  with no $N=2$ supersymmetry sector \cite{Antoniadis}.} The latter is
given by the expression
\begin{eqnarray}
{\Lag}_{\rm GS}&=&{\superint}EV{V_{\rm GS}},\\
{V_{\rm GS}}&=& b{\sum_I}g^{I} + {\sum_{A}}p_{A}
e^{{\sum_I}{q_{I}^{A}}{g^{I}}} \left|{\Phi^{A}}\right|^{2} +
  \order\( \left|{\Phi^{A}}\right|^{4} \),
\label{eq:GSterm}
\end{eqnarray}
where $b\equiv C_{E_8}/8{\pi}^2 \approx 0.38$ is
proportional to the beta-function coefficient for the group $E_8$ and
the coefficients $p_A$ are as yet undetermined.

As for the operators ${\Lag}_{\rm a}$ in~(\ref{eq:completeLag}),
their rather involved form in curved superspace was worked out
in~\cite{gauginomass} and will not be repeated here. Their importance
for this work lies in their contributions to the supersymmetry-breaking gaugino 
masses at the condensation scale arising from the superconformal anomaly -- 
a contribution that was recently emphasized by a number of
authors~\cite{anomaly}. We will return to these in Section~\ref{sec:softterms}.

\subsection{Condensation and Dilaton Stabilization}
\label{sec:condensation}
The Lagrangian in~(\ref{eq:completeLag}) can be expanded into component
form using the standard techniques of the K\"ahler superspace
formalism of supergravity \cite{BGG}. In reference~\cite{ModInv}
the bosonic part of the Lagrangian relevant to dilaton stabilization
and gaugino condensation was presented and the equations of motion for 
the nonpropagating fields were solved. In particular, the equations of
motion for the auxiliary fields of the condensates $U^{a}$ give
\begin{equation}
{{\rho_a}^2}=e^{-2{\frac{b'_a}{b_a}}}e^{K}e^{-\frac{\(1+f\)}{{b_a}\dilaton}}e^{-\frac{b}{b_a} 
  {\sum_I}g^{I}}{\prod_I}\left|{\eta}\(t^{I}\)\right|^{\frac{4\(b-b_{a}\)}{b_a}}
{\prod_{\alpha}}\left|\baal/4c_{\alpha}\right|^{-2 {\frac{b_{a}^{\alpha}}{b_a}}},
\label{eq:cond1}
\end{equation}
where $t_{I}\equiv T_{I}\lowest$ and $u_{a} = U_{a}\lowest \equiv
{\rho}_{a}e^{i{\omega}_a}$.

Upon substituting for the gauge coupling via the relation 
${g^2}/{2}= {\dilaton}/{\(1+f\(\dilaton\)\)}$ we recognize the
expected one-instanton form for gaugino
condensation. Expression~(\ref{eq:cond1}) encodes more information,
however, than simply the one-loop running of the gauge coupling. In
\cite{match} the loop corrections to the gauge coupling constants were 
computed using a manifestly supersymmetric Pauli-Villars
regularization. The (moduli independent) corrections were identified
with the renormalization group invariant \cite{Russians}
\begin{equation}
{\delta}_a =\frac{1}{g_{a}^{2}\(\mu\)} - 
\frac{3b_a}{2}\ln{\mu^2}
+\frac{2C_a}{16{\pi}^2}\ln{g_{a}^{2}\(\mu\)} + 
\frac{2}{16{\pi}^2}\sum_{a}C_{a}^{A}\ln{Z_{a}^{A}\(\mu\)}.
\label{eq:RGinvariant}
\end{equation}
Using the above expression it is possible to solve for the scale at
which the $1/g^{2}(\mu)$ term becomes negligible relative to the 
$\ln{g^{2}(\mu)}$ term -- effectively looking for the ``all loop'' Landau
  pole for the coupling constant. This scale is related to the string scale by the relation
\begin{equation}
{\mu_L}^2 \sim {\mu_{\rm str}}^2 e^{-\frac{2}{3b_{a}g_{a}^{2}\(\mu\)}}
\prod_{A}\[Z_{a}^{A}\(\mu_{\rm str}\)/Z_{a}^{A}\(\mu_{L}\)\]^
{\frac{C_{a}^{A}}{12{{\pi}^2}b_{a}}}.
\label{eq:Landau}
\end{equation}
Now comparing the effective Lagrangian given in Section~\ref{sec:Lag}
with the field theory loop calculation given in \cite{match} shows
that the two agree provided we identify the wave function
renormalization coefficients $Z_{a}^{A}$ with the quantity 
$|4W_{\alpha}/\baal|^{2}$. This is precisely what is needed to
produce the final product in the condensate expression given
in~(\ref{eq:cond1}), indicating that the condensation scale represents 
the scale at which the coupling becomes strong as would be computed
using the so-called ``exact'' beta-function. 

Note that this final
factor introduces the unknown Yukawa coefficients $c_{\alpha}$ into
the scale of supersymmetry breaking. Such dependence of the gaugino
condensate on the parameters of the superpotential is not unexpected,
and has in fact been demonstrated in the case of supersymmetric QCD as well as
certain models of supersymmetric Yang-Mills theories coupled to chiral 
matter \cite{Amati}.
This last Yukawa-related factor has the
virtue of allowing two different hidden sector configurations which
result in the same beta-function to condense at widely different
scales.

In order to go further and make quantitative statements about the scale of
gaugino condensation (and hence supersymmetry breaking) it is
necessary to specify some form for the nonperturbative effects
represented by the functions $f$ and $g$. The parameterization adopted 
in \cite{susybreak} was originally motivated by Shenker~\cite{Shenker}
and was of the form $\exp{\(-1/{g_{\rm str}}\)}$ where $g_{\rm str}$ is the string
coupling constant. A consensus seems to be forming~\cite{nonpert} around this
characterization for string nonperturbative effects and the function
$f\(V\)$ in~(\ref{eq:LKE}) will be taken to be of the form
\begin{equation}
f\(V\)=\[{A_0}+{A_1}/\sqrt{V}\]e^{-B/\sqrt{V}},
\label{eq:nonpertf}
\end{equation}
which was shown \cite{susybreak} to allow dilaton stabilization at
weak to moderate string coupling with parameters that are all of
$\order \(1\)$. The benefits of invoking string-inspired
nonperturbative effects of the form of~(\ref{eq:nonpertf}) have
recently been explored by others in the literature \cite{casas}.

The scalar potential for the moduli $t_{I}$ is
minimzed at the self-dual points $\lang t_{I} \rang =1$ or $\lang
t_{I} \rang = \exp{\(i\pi/6\)}$, where the corresponding F-components $F_{I}$
of the chiral superfields $T^{I}$ vanish. At these points the dilaton
potential is given by
\begin{equation}
V\(\dilaton\)=\frac{1}{16{\dilaton}^2}\(1+\dilaton \frac{\diff
  g}{\diff \dilaton}\) \left|{\sum_a}\(1+{b_a}\dilaton\)
  {u_a}\right|^{2}- \frac{3}{16}\left|{\sum_a} {b_a}{u_a}\right|^2.
\label{eq:dilpot}
\end{equation}
As an example, the potential~(\ref{eq:dilpot}) can be minimized with
vanishing cosmological constant and $\alpha_{\rm str}=0.04$ for
$A_{0}=3.25, A_{1}=-1.70$ and $B=0.4$ in expression~(\ref{eq:nonpertf}).

\section{Phenomenological Implications}
\label{sec:phenom}
\subsection{Scale of Supersymmetry Breaking}
\label{sec:susybreaking}
With the adoption of~(\ref{eq:nonpertf}) the scale of gaugino
condensation can be obtained once the following are specified: ({\bf 1}) the
condensing subgroup(s) of the {\mbox original} hidden sector gauge group
$E_8$, ({\bf 2}) the representations of the matter fields charged under the condensing
subgroup(s), ({\bf 3}) the Yukawa coefficients in the superpotential for the
hidden sector matter fields and ({\bf 4}) the value of the string coupling
constant at the compactification scale, which in turn determines the 
coefficients in~(\ref{eq:nonpertf}) necessary to minimize the scalar potential~(\ref{eq:dilpot}).

A great deal of simplification in the above parameter space can be
obtained by making the ansatz that all of the matter in the hidden
sector which transforms under a given subgroup ${\cal G}_a$ is of the
same representation, such as the fundamental representation. Then the
sum of the coefficients $\baal$ over the number of condensate fields
labeled by $\alpha$ $\(\alpha = 1,\ldots,N_c\)$, can be replaced by one
effective variable
\begin{eqnarray}
{\sum_{\alpha}}\baal\longrightarrow\baaleff & 
\baaleff={N_c}b_{a}^{\rm rep}.
\label{eq:baaleff}
\end{eqnarray}
In the above equation $b_{a}^{\rm rep}$ is proportional to the
quadratic Casimir operator for the matter fields in the common
representation and the number of condensates, $N_c$, can range from
zero to a maximum value determined by the condition that the gauge
group presumed to be condensing must remain asymptotically free. The
redefinition in~(\ref{eq:baaleff}) essentially takes the coefficients
$\baal$, which we are free to choose in our effective Lagrangian up to 
the conditions given in~(\ref{eq:coeff}), and assigns the same value to
each condensate.

The variable $\baal$ can then be eliminated in~(\ref{eq:cond1}) in favor
of $\baaleff$ provided the simultaneous redefinition
$c_{\alpha}\longrightarrow\(c_{\alpha}\)_{\rm eff}$
is made so as to keep the product in~(\ref{eq:cond1}) invariant:
\begin{equation}
\lang\rho_{+}^{2}\rang\sim {\left|\frac{\baaleff}{4\caleff}\right|}^{-2{\baaleff}/b_{a}}.
\label{eq:cond2}
\end{equation}
With the assumption of universal representations for the matter
fields, this implies
\begin{equation}
\caleff \equiv {N_c}\({\prod_{\alpha=1}^{N_c}}c_{\alpha}\)^\frac{1}{N_c}
\label{eq:caleff}
\end{equation}
which we assume to be an $\order\(1\)$ number, if not slightly smaller.

From a determination of the condensate value $\rho$
using~(\ref{eq:cond1}) the supersymmetry-breaking scale can be found
by solving for the gravitino mass, given by
\begin{equation}
M_{3/2}=\frac{1}{3}\lang\left|M\right|\rang=\frac{1}{4}\lang
\left|{\sum_a}{b_a}{u_a} \right|\rang.
\label{eq:gravmasssum}
\end{equation}
In~\cite{ModInv} it was shown that in the case of multiple gaugino
condensates the scale of supersymmetry breaking was governed by the
condensate with the largest one-loop beta-function coefficient. Hence
in the following it is sufficient to consider the case with just one
condensate with beta-function coefficient denoted $b_+$:
\begin{equation}
M_{3/2}=\frac{1}{4}{b_+}\lang\left|u_+\right|\rang.
\label{eq:gravmass}
\end{equation}
As an illustration of this point, the gravitino mass for the case of
pure supersymmetric
Yang-Mills $SU(5)$ condensation (no hidden
sector matter fields) would be $4000$ GeV. The addition of an
additional condensation of pure supersymmetric Yang-Mills $SU(4)$
gauginos would only add an additional $0.004$ GeV to the mass.

\begin{figure}[t]
%    \begin{center}
\centerline{
       \psfig{file=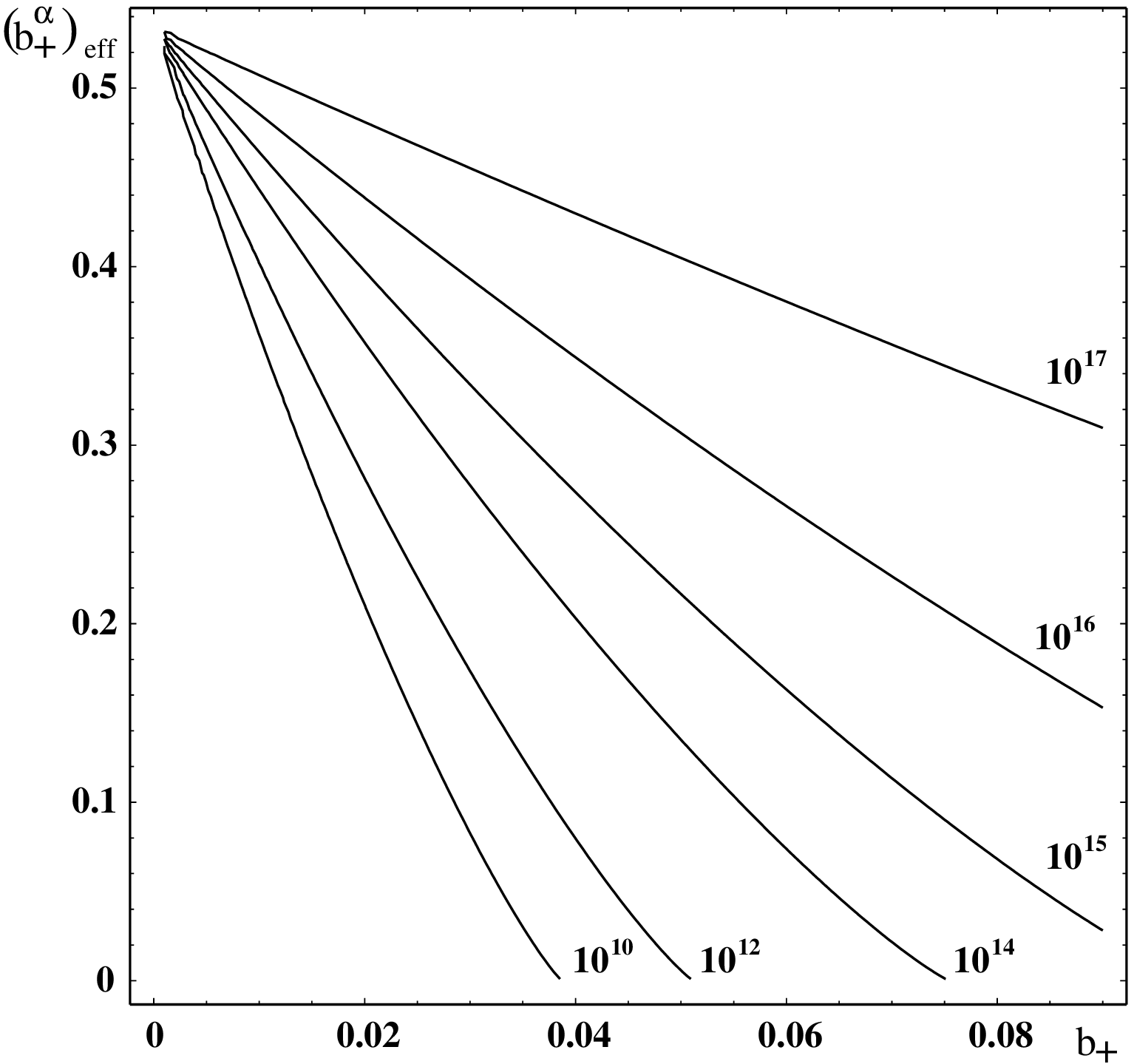,width=0.5\textwidth}
       \psfig{file=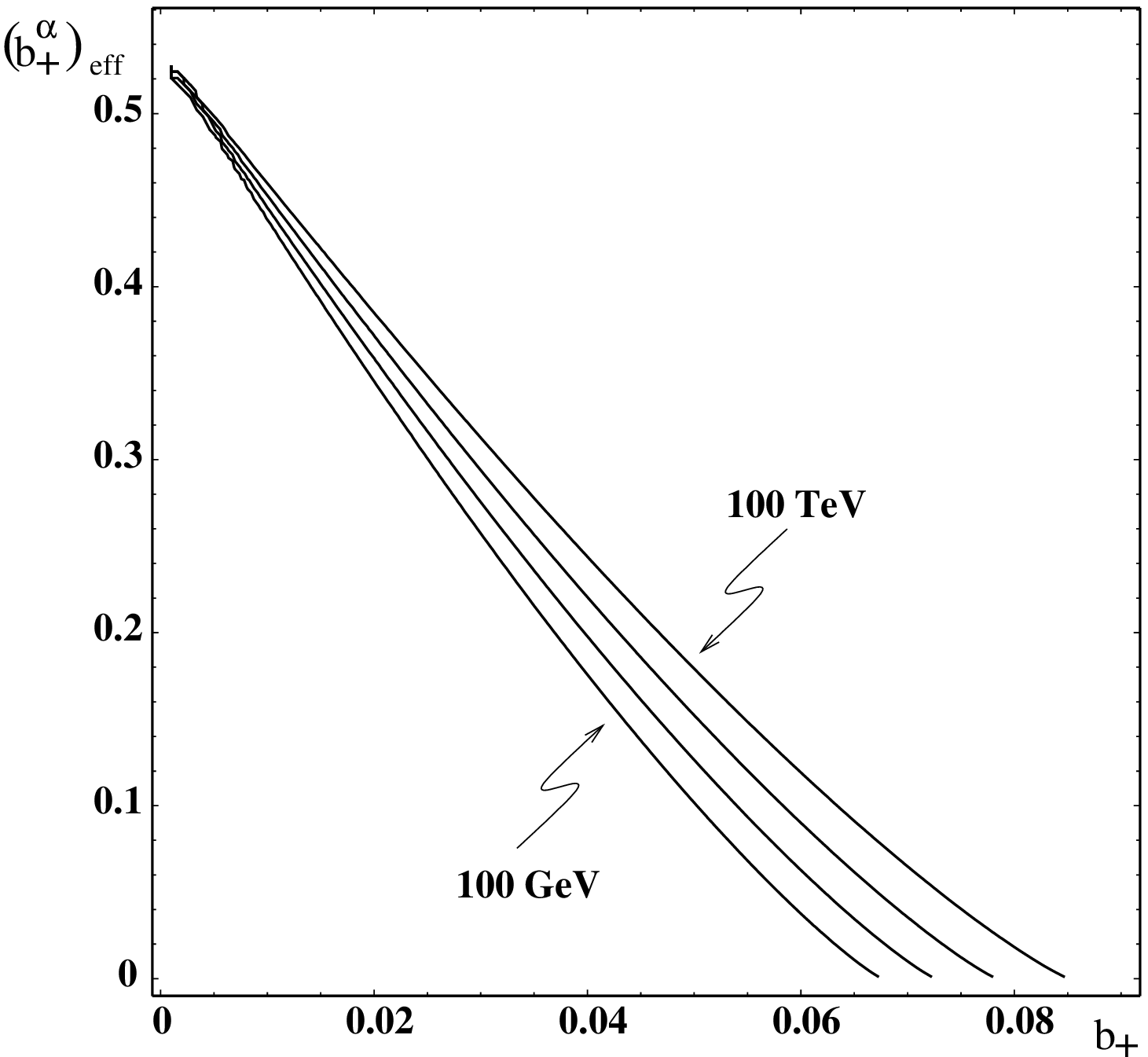,width=0.5\textwidth}}
          \caption{{\footnotesize {\bf Condensation Scale and
                Gravitino Mass}. Contours give the
            scale of gaugino condensation in GeV in the left panel and gravitino masses
            of $10^2$ through $10^5$~GeV in the right panel for $\caleff=3$.}}
        \label{fig:conscale}
%    \end{center}
\end{figure}

Now for given values of $\caleff$ and $g_{\rm str}$ the condensation
scale
\begin{equation}
\Lambda_{\rm cond}=\(M_{\rm Planck}\)\lang {\rho}^{2}_{+}\rang^{1/6}
\label{eq:conscale}
\end{equation}
and gravitino mass~(\ref{eq:gravmass}) can be plotted in the $\lbr b_{+},
\bpaleff\rbr$ plane. The sharp
variation of the condensate value with the parameters of the theory,
as anticipated by the functional form in~(\ref{eq:cond1}), is
apparent in the contour plot of Figure~\ref{fig:conscale}. 

The dependence of the gravitino mass on the group theory
parameters is even more profound. Figure~\ref{fig:conscale} gives
contours for the
gravitino mass between $100$ GeV and $100$ TeV. Clearly, the region of 
parameter space for which a phenomenologically preferred value of the
supersymmetry-breaking scale occurs is a rather limited slice of the
entire space available.

\begin{figure}[t]
%    \begin{center}
\centerline{
       \psfig{file=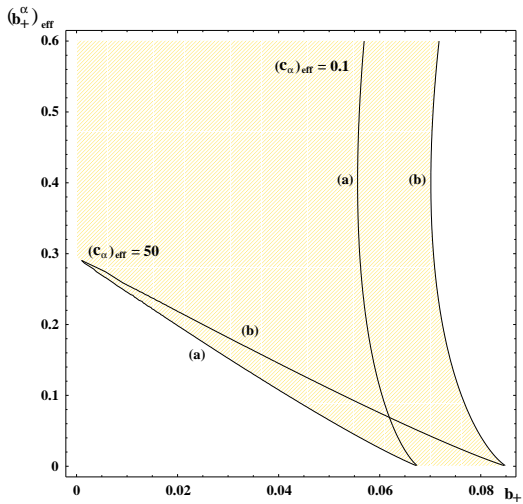,width=0.5\textwidth}}
          \caption{{\footnotesize {\bf Gravitino Mass Regions as a Function of Yukawa 
              Parameter}. Gravitino mass contours for (a) 100~GeV and
            (b) 10~TeV are shown for $\caleff=50$ and $\caleff=0.1$
            with $\alpha_{\rm str}=0.04$. The
            region between the two sets of curves can be considered
            roughly the region of phenomenological viability.}}
        \label{fig:gravmassvaryc}
%    \end{center}
\end{figure}

The variation of the gravitino mass as a function
of the Yukawa parameters $c_{\alpha}$ is shown in
Figure~\ref{fig:gravmassvaryc}. On the
horizontal axis there are no matter condensates ($\baal=0,\ \forall 
\alpha$) so there is no dependence on the variable $\caleff$. 
For values of $\caleff < 0.1$ the contours of gravitino mass in the
TeV region lie beyond the limiting value of $b_{+} \simeq 0.09$ 
and are thus in a region of parameter space which is inaccessible to a model in 
which the unified coupling at the string scale is $\alpha_{\rm
  str}=0.04$ or larger. For very large values of the effective Yukawa
parameter the gravitino mass contours approach an asymptotic value
very close to the case shown here for $\caleff = 50$. We might
therefore consider the shaded region between the two sets of contours
as roughly the maximal region of viable parameter space for a given
value of the unified coupling at the string scale.

\subsection{Implications for the Hidden Sector}
\label{sec:implication}
Having examined some of the universal constraints placed on any
string-derived model proposing to describe low energy physics in
Section~\ref{sec:susybreaking} it is natural to ask 
whether the region of phenomenological viability (roughly the shaded
area in Figure~\ref{fig:gravmassvaryc}) can be used to constrain the
matter content of the hidden sector.

Upon orbifold compactification the $E_8$ gauge group of the hidden
sector is presumed to break to some subgroup(s) of $E_8$ and the set of
all such possible breakings has been computed for $Z_N$
orbifolds~\cite{KEKtable}. Under the working assumption that only the
subgroup with the largest beta-function coefficient enters into the
low-energy phenomenology, there are then a finite number of possible
groups to consider: 
\begin{equation}
\lbr \begin{array}{l}
E_{7}, E_{6} \\
SO\(16\), SO\(14\), SO\(12\), SO\(10\), SO\(8\) \\
SU\(9\), SU\(8\), SU\(7\), SU\(6\), SU\(5\), SU\(4\), SU\(3\)
\end{array} \right.
\label{eq:groups}
\end{equation}

For each of the above gauge groups equations~(\ref{eq:coeff})
and~(\ref{eq:betafunc}) define a 
line in the $\lbr b_{+},\bpaleff\rbr$ plane. These lines will all be
parallel to one another with horizontal intercepts at the
beta-function coefficient for a pure Yang-Mills theory.
The vertical intercept will then indicate the amount of matter
required to prevent the group from being asymptotically free, thereby
eliminating it as a candidate source for the supersymmetry breaking described 
in Section~\ref{sec:susybreaking}.

\begin{figure}[t]
%    \begin{center}
\centerline{
       \psfig{file=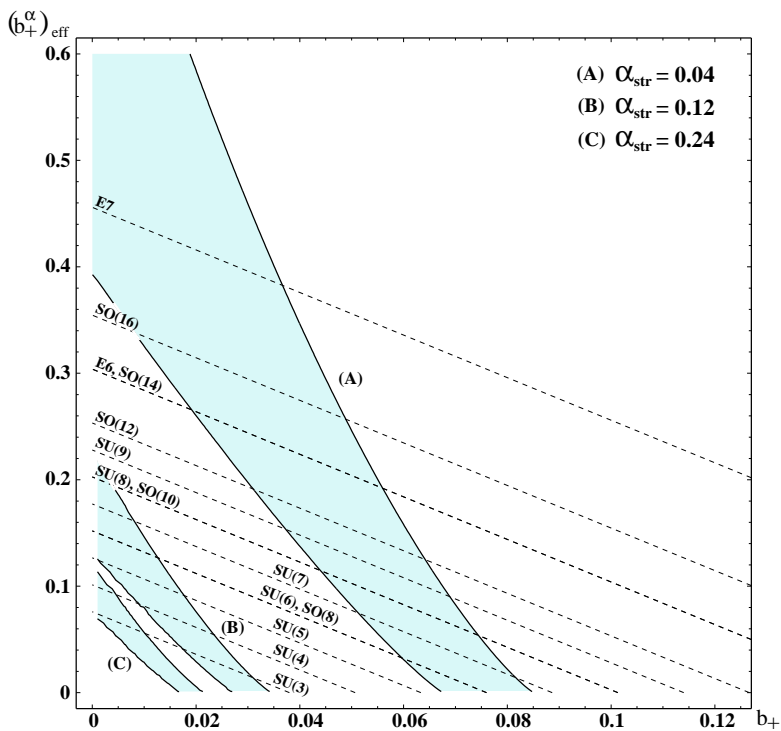,width=0.75\textwidth}}
          \caption{{\footnotesize {\bf Constraints on the Hidden
                Sector}. The shaded regions give three different
              ``viable'' regions depending on the value of the unified
              coupling strength at the string scale. The upper limit
              in each case represents a 10~TeV gravitino mass contour
              with $\caleff=1$, while the lower bound represents a
              100~GeV gravitino mass contour with $\caleff=10$.}}
        \label{fig:bestplot}
%    \end{center}
\end{figure}

In Figure~\ref{fig:bestplot} we have overlaid these gauge lines on a
plot similar to Figure~\ref{fig:gravmassvaryc}. We restrict the Yukawa 
couplings of the hidden sector to the more reasonable range of $1
\leq \caleff \leq 10$ and give three different values of the string
coupling at the string scale. The choice of string coupling constant is made when
specifying the
boundary conditions for solving the dilaton scalar potential, as
described in Section~\ref{sec:condensation}. Changing this boundary
condition will affect the scale of gaugino condensation through
equation~(\ref{eq:cond1}), altering the supersymmetry-breaking scale
for a fixed point in the $\lbr b_{+},\bpaleff\rbr$
plane. Demanding larger values of $g_{\rm str}$ will result in the
shifting of the contours of fixed
gravitino mass towards the origin, as in Figure~\ref{fig:bestplot}.
Such large values of $\alpha_{\rm str}$ have recently been invoked as
part of a mechanism for stabilizing the dilaton
and/or as a consequence of reconciling the apparent scale of gauge
unification in the minimal supersymmetric standard model (MSSM) with
the scale predicted by string theory~\cite{gauge}.
We will return to such issues in Section~\ref{sec:gaugeunify}.

A typical matter configuration would be represented in
Figure~\ref{fig:bestplot} by a point on one of the gauge group
lines. As each field adds a discrete amount to $\baaleff$ and the
fields must come in gauge-invariant multiples, the set of all such
possible hidden sector configurations is necessarily a finite
one.\footnote{For example, one cannot obtain values of $b_{+}$
  arbitrarily close to zero in practical model building.} The
number of possible configurations consistent with a given choice of
$\lbr \alpha_{\rm str}, \caleff \rbr$ and supersymmetry-breaking scale $M_{3/2}$
is quite restricted. For example, 
Figure~\ref{fig:bestplot} immediately rules out hidden sector gauge groups smaller
than SU(6) for weak coupling at the string scale $\(g^{2}_{\rm str} \simeq 
0.5\)$. Furthermore, even moderately larger values of the string
coupling at unification become increasingly difficult to obtain as it
is necessary to postulate a hidden sector with very small gauge group
and particular combinations of matter to force the beta-function
coefficient to small values.

\section{Constraints from the Low-Energy Spectrum}
\label{sec:RGE}
\subsection{Soft Supersymmetry-Breaking Terms}
\label{sec:softterms}
Simply requiring that the scale of supersymmetry breaking be in a
reasonable range of energy values ({\em i.e.} within an order of magnitude
of 1 TeV) can put significant constraints on the dynamics of the
hidden sector. Requiring further that the {\em pattern} of
supersymmetry breaking be consistent with observed electroweak
symmetry breaking and direct experimental bounds on superpartner
masses can restrict the parameter space even more.

The pattern of supersymmetry breaking is determined by the appearance
of soft scalar masses, gaugino masses and trilinear couplings at the
condensation scale. The gaugino masses in the one-condensate
approximation, including the contribution from the quantum effects of
light fields arising at one loop from the superconformal anomaly, are
given by~\cite{gauginomass}
\begin{equation}
m_{{\lambda}_a}|_{\mu = \Lambda_{\rm cond}} =
-\frac{g^{2}_{a}\(\mu\)}{2} \[\frac{3{b_+}\(1+{b'_a}\ell\)}{1+{b_+}\ell}
-3{b_a}
+{\sum_A}\frac{{C_{a}^{A}}{p_A}\(1+{b_+}\ell\)}{4{{\pi}^2}{b_+}\(1+{p_A}\ell\)} 
\] M_{3/2}.
\label{eq:gaugino}
\end{equation}

The incorporation of scalar masses and trilinear terms in the scalar
potential for observable sector matter fields ${\Phi^A}$ depends on
the form of the K\"ahler potential and the nature of the couplings of
observable sector matter fields to the Green-Schwarz counterterm. 
Adopting the K\"ahler potential assumed in~(\ref{eq:Kahlerpot}) and the
counterterm of~(\ref{eq:GSterm}), the scalar
masses are given in the one-condensate approximation by
\begin{equation}
m^{2}_{A} = \frac{1}{16} \lang {\rho_{+}^{2}} \frac{\(p_{A} -
  b_{+}\)^{2}}{\(1+{p_A}\dilaton\)^{2}} \rang,
\label{eq:scalarmass}
\end{equation}
and the trilinear ``A-terms'' in the scalar potential are given by
\begin{equation}
V_{A}\(\phi\)=Ae^{K/2}W\(\phi\)
\end{equation}
with
\begin{equation}
A=\frac{3}{4} \lang {\bar{u}}_{+}\rang
\lang \frac{\(p_{A}-b_{+}\)}{\(1+{p_A}\dilaton\)}+
\frac{b_{+}}{\(1+{b_+}\dilaton\)} \rang.
\label{eq:Aterms}
\end{equation}

As noted in~\cite{susybreak}, the fact that~(\ref{eq:scalarmass})
and~(\ref{eq:Aterms}) are 
independent of the modular weights $q_{I}^{A}$ of the individual
observable sector fields is the result of the vanishing of the
auxiliary fields $F^{I}$ in the vacuum. This is a manifestation of the
so-called ``dilaton dominated'' scenario of supersymmetry breaking~\cite{dildom} for 
which flavor-changing neutral currents might be naturally
suppressed. For this to indeed occur, however, it is also necessary to 
make the assumption that the couplings $p_A$ are the same for the
first and second generations of matter.

To analyze the low-energy particle spectrum it is necessary to choose
a value of $p_A$ for each generation of matter fields. If the
Green-Schwarz term~(\ref{eq:GSterm}) is independent of the ${\Phi^A}$
so that $p_{A}=0$, then from~(\ref{eq:scalarmass}) $m_{A}=M_{3/2}$. We 
will call such a generation ``light.'' On
the other hand, it was postulated in~\cite{susybreak} that the
Green-Schwarz term may well depend only on the combination
$T^{I}+{\bar{T}}^{I}-{\sum_A} \left|{\Phi}_{I}^{A}\right|^{2}$, where
${\Phi}_{I}^{A}$ represents untwisted matter fields. Then for these
multiplets $p_{A}=b$ and the scalar masses for these fields are in
general an order of magnitude greater than the gravitino mass. We will 
call these generations ``heavy.''

The scalar masses~(\ref{eq:scalarmass}) and A-terms~(\ref{eq:Aterms})
given above do not include the contributions proportional to the matter
field wave-function renormalization coefficients arising from the
superconformal anomaly (the analog to the gaugino mass
terms studied in \cite{gauginomass} and included
in~(\ref{eq:gaugino})). A systematic treatment of these
contributions to the soft-breaking terms is currently underway
\cite{inprogress}, but their general size is comparable to the
gaugino masses. In the following it has been checked that varying the
initial soft terms by arbitrary amounts of this size has a negligible impact on
the conclusions we report here.

Before giving the results of a numerical analysis using the
renormalization group equations (RGEs) with the boundary conditions
determined by equations~(\ref{eq:gaugino}),~(\ref{eq:scalarmass})
and~(\ref{eq:Aterms}), it is worthwhile looking at what patterns of
symmetry breaking are expected for various choices of the
parameter $p_A$ in the context of the MSSM. For any generation with non-negligible Yukawa
couplings a good indicator that the stable minimum of the scalar potential
will yield correct electroweak symmetry breaking is the relation
\begin{equation}
\left|A_t\right|^{2} \leq 3\(m_{Q}^{2}+m_{U}^{2}+m_{H_u}^{2}\).
\label{eq:CCB}
\end{equation}
When this bound is not satisfied it is typical to develop minima away
from the electroweak symmetry breaking point in a direction in which
one of the scalar masses carrying electric or color charge becomes negative.
For any heavy matter generation with a non-negligible coupling to a heavy Higgs field 
($p_{A}=b$) equation~(\ref{eq:Aterms}) yields 
$A \approx 3m_{A}$ and so~(\ref{eq:CCB}) is already nearly saturated
at the condensation scale.

Another key factor in preventing dangerous color and charge-breaking
minima is the ratio of scalar masses to gaugino masses and the degree
of {\mbox splitting} between any light and heavy matter
generations. In this model, both of the hierarchies, $m_{A}^{\rm
  light}/m_{\lambda}$ and $m_{A}^{\rm
  heavy}/m_{A}^{\rm light}$, will turn out to be $\order \(10\)$. This 
pattern of soft supersymmetry-breaking masses has been shown
\cite{CCB} to lie on 
the boundary of the region in MSSM parameter space for which light
squark masses tend to be driven negative by two-loop effects arising
from the heavier squarks. All of the above considerations suggest that 
compactification scenarios in which the observable sector matter
fields couple universally to the Green-Schwarz counterterm with $p_{A} 
=b$ may have trouble reproducing the correct pattern of low-energy symmetry breaking.

\subsection{RGE Viability Analysis Within the MSSM}
\label{sec:viable}
To determine what region of parameter space in the  $\lbr
b_{+},\bpaleff\rbr$ plane is consistent with current experimental data 
it is necessary to run the soft supersymmetry-breaking parameters of
equations~(\ref{eq:gaugino}),~(\ref{eq:scalarmass})
and~(\ref{eq:Aterms}) from the condensation scale to the electroweak
scale using the renormalization group equations. For this purpose we
take the MSSM superpotential and matter content for the observable
sector, keeping only the top, bottom and tau Yukawa couplings. In
order to capture the significant two loop contributions to gaugino
masses and scalar masses these parameters are run at two loops,
while the other parameters are evolved using the one-loop RGEs. The
equations used are in the $\overline{DR}$ scheme and can be found in \cite{RGE}.
The RGE analysis was performed on four different scenarios:
\begin{itemize}
\item {\bf Scenario A}: All three generations light.
\item {\bf Scenario B}: Third generation light, first and second generations
  heavy.
\item {\bf Scenario C}: All three generations heavy.
\item {\bf Scenario D}: All matter heavy except for the two Higgs
  doublets which remain light ($p_{A}=0$). 
\end{itemize}

To protect against unwanted flavor changing neutral currents we have
chosen the Green-Schwarz coefficients $p_{A}$ to be universal
throughout each matter generation. While our scalars will turn out to
be heavy enough that small deviations from universality (such as those 
arising from the superconformal anomaly discussed above) will not be
problematic, the large hierarchies controlled by the values of the
$p_{A}$ would be untenable. The Higgs fields will be taken to couple
to the Green-Schwarz counterterm identically to the third generation
of matter, as we keep only the third generation Yukawa couplings in the 
MSSM superpotential. In Scenario D we relax this assumption.

In the boundary values of~(\ref{eq:gaugino}),~(\ref{eq:scalarmass})
and~(\ref{eq:Aterms}) the values of $\caleff$ and $\baaleff$ appear
only indirectly through the determination of the value of the
condensate $\lang \rho_{+}^{2} \rang$. It is thus convenient to cast all
soft supersymmetry-breaking parameters in terms of the values of
$b_{+}$ and $M_{3/2}$ using equation~(\ref{eq:gravmass}). While the
gravitino mass itself is not strictly independent of $b_{+}$, it is
clear from Figure~\ref{fig:gravmassvaryc} that we are
guaranteed of finding a reasonable set of values for $\lbr \caleff
,\bpaleff\rbr$ consistent with the choice of $b_{+}$ and $M_{3/2}$
provided we scan only over values $b_{+} \leq 0.1$ for weak string
coupling. This transformation of variables allows the slice of
parameter space represented by the contours of
Figure~\ref{fig:bestplot} to be recast as a
two-dimensional plane for a given value of $\tan{\beta}$ and ${\rm
  sgn}(\mu)$. The condensation scale (the scale at which the RG-running
begins) is also a function of the gravitino mass in this framework,
found by inverting equation~(\ref{eq:gravmass}).

Having chosen a set of input parameters $\lbr b_{+}, M_{3/2},
\tan{\beta}, {\rm sgn}(\mu) \rbr$ for a particular scenario, the model 
parameters are run from the condensation scale $\Lambda_{\rm cond}$
given by~(\ref{eq:conscale}) to
the electroweak scale $\Lambda_{\rm EW} = M_{Z}$, decoupling the
scalar particles at a scale approximated by $\Lambda_{\rm scalar} =
m_{A}$. While treating all superpartners with a common scale
sacrifices precision for expediency, the results presented below are
meant to be a first survey of the phenomenology of this class of
models. 

At the electroweak scale the one-loop corrected effective
potential $V_{\rm 1-loop}=V_{\rm tree} + \Delta V_{\rm rad}$ is
computed and the effective mu-term $\bar{\mu}$ is calculated
\begin{equation}
{\bar{\mu}}^{2}=\frac{\(m_{H_d}^{2}+\delta m_{H_d}^{2}\) -
  \(m_{H_u}^{2}+\delta m_{H_u}^{2}\) \tan{\beta}}{\tan^{2}{\beta}-1}
-\frac{1}{2} M_{Z}^{2}.
\label{eq:radmuterm}
\end{equation}
In equation~(\ref{eq:radmuterm}) the quantities $\delta m_{H_u}$ and $\delta m_{H_d}$ are
the second derivatives of the radiative corrections $ \Delta V_{\rm
  rad}$ with respect to the up-type and down-type Higgs scalar fields, 
respectively. These corrections include the effects of all
third-generation particles. If the right hand side of equation~(\ref{eq:radmuterm})
is positive then there exists some initial value of $\mu$ at the
condensation {\mbox scale} which results in correct electroweak symmetry
breaking with $M_{Z} = 91.187$~GeV~\cite{radcorr}.\footnote{Note that we do not 
try to specify the origin of this mu-term (nor its associated
``B-term'') and merely leave them as free parameters in the theory --
ultimately determined by the requirement that the Z-boson receive the
correct mass.}

A set of input parameters will then be considered viable if at the
electroweak scale the one-loop corrected mu-term $\bar{\mu}^2$ is
positive, the Higgs potential is bounded from below, all matter fields have
positive scalar mass-squareds and the spectrum of physical masses for the
superpartners and Higgs fields satisfy the selection criteria given
in Table~\ref{tbl:massbounds}.\footnote{Though the inclusive branching ratio for $b
  \rightarrow s\gamma$ decays was not used as a criterion, an {\em a
    posteriori} check of the
  region of the parameter space where this class of models wants to
  live -- namely relatively low $\tan{\beta}$ and gaugino masses with high scalar masses
  -- indicates no reason to fear a conflict with the bounds from CLEO
  except possibly in the case ${\rm sgn}(\mu)=-1$ for Scenario~D \cite{bsgamma}.}

\begin{table}[htb]
\caption{Superpartner and Higgs mass constraints imposed~\cite{bounds}.}
{\begin{center}
\begin{tabular}{|llrcr|} \cline{1-5}
Gluino Mass & \hspace{2mm} & $m_{\tilde{g}}$&$>$&$175$ GeV \\
Lightest Neutralino Mass & \hspace{2mm} & $m_{\tilde{N_1}}$&$>$&$15$ GeV \\
Lightest Chargino Mass & \hspace{2mm} & $m_{\tilde{\chi_{1}}^{\pm}}$&$>$&$70$ GeV \\
Squark Masses & \hspace{2mm} & $m_{\tilde{q}}$&$>$&$175$ GeV \\
Slepton Masses & \hspace{2mm} & $m_{\tilde{l}}$&$>$&$50$ GeV \\
Light Higgs Mass & \hspace{2mm} & $m_{h}$&$>$&$80$ GeV \\
Pseudoscalar Higgs Mass & \hspace{2mm} & $m_{A}$&$>$&$65$ GeV \\
\cline{1-5}
\end{tabular}
\end{center}}
\label{tbl:massbounds}
\end{table}

\begin{figure}[t]
%    \begin{center}
\centerline{
       \psfig{file=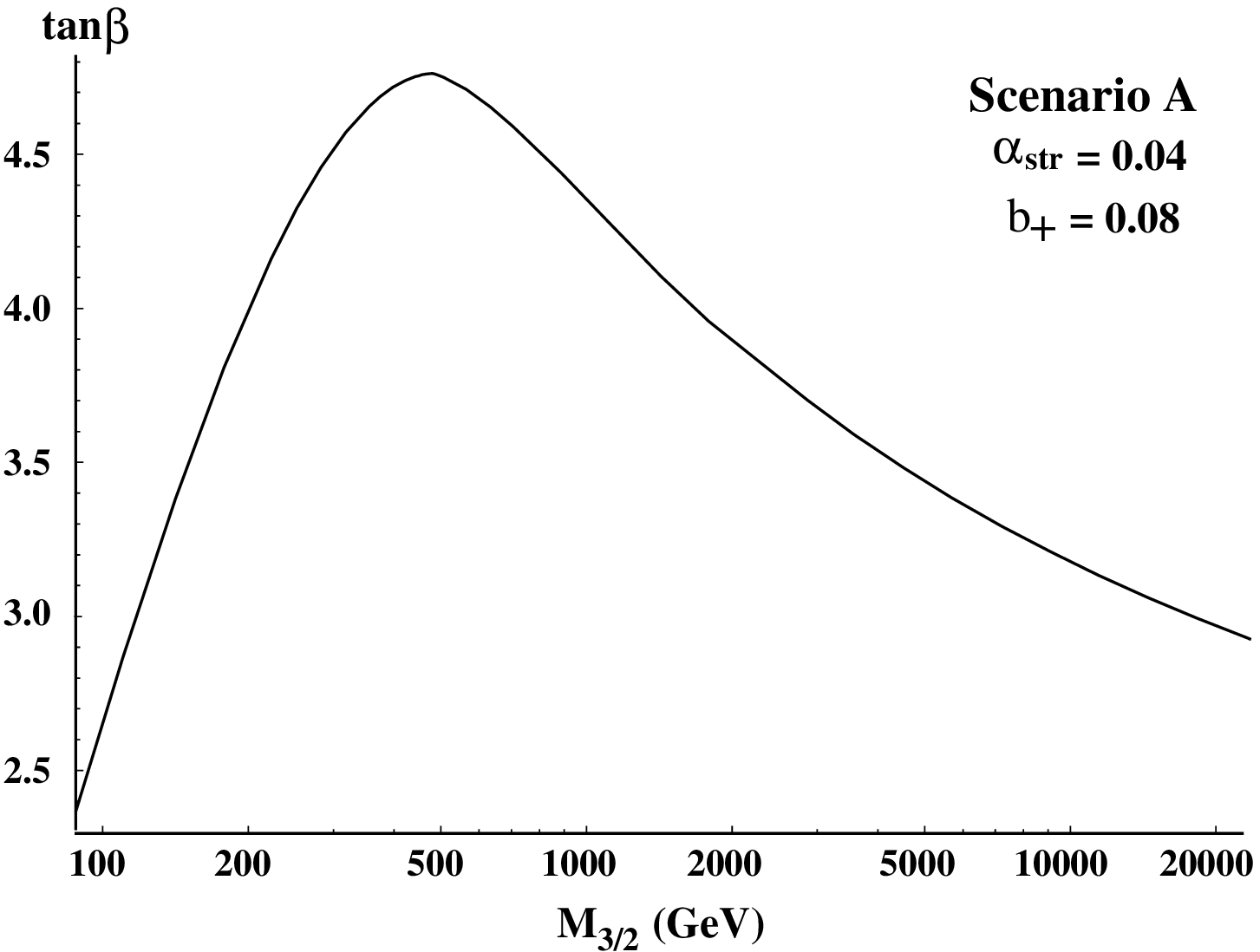,width=0.5\textwidth}
       \psfig{file=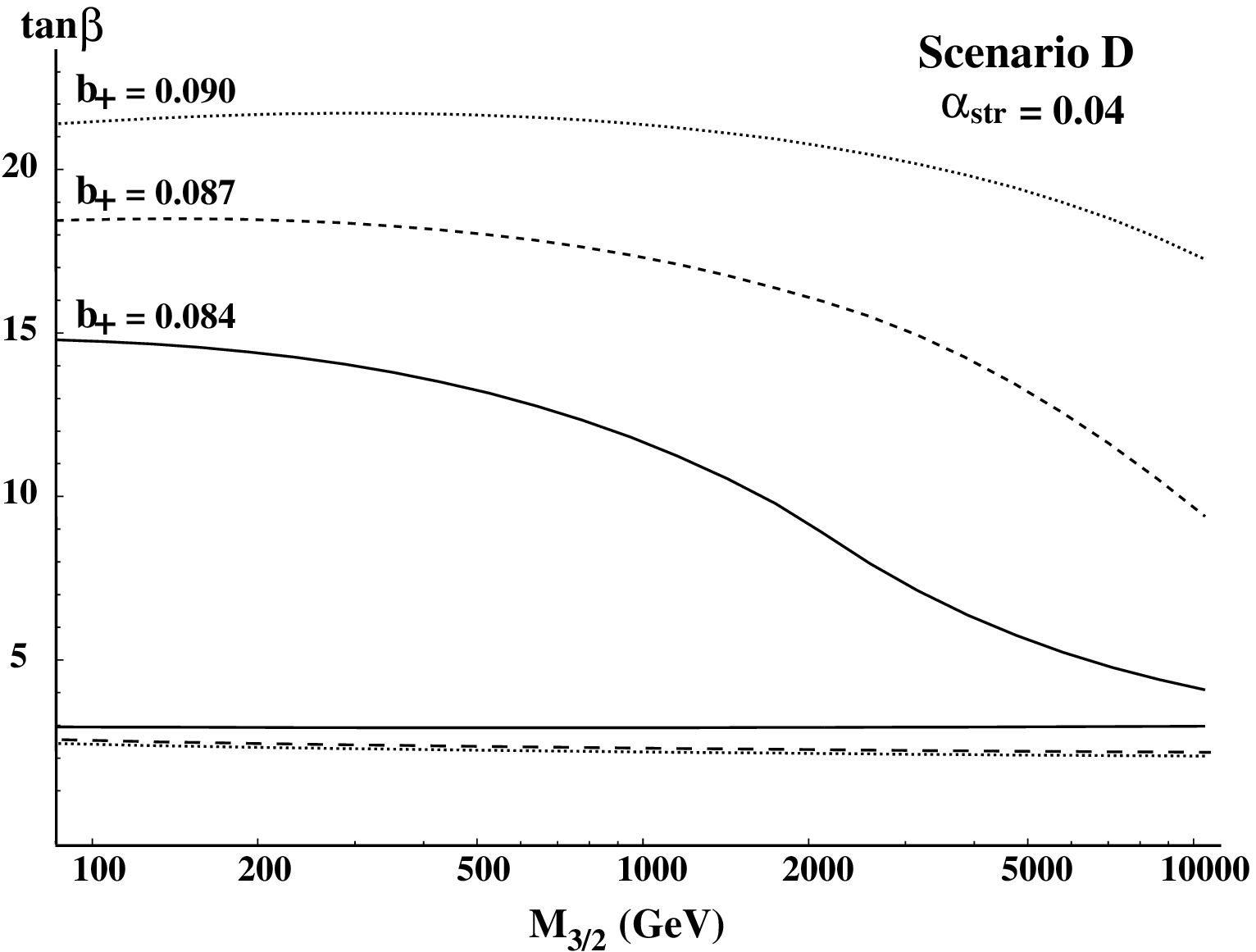,width=0.5\textwidth}}
          \caption{{\footnotesize {\bf Region with Correct Symmetry
                Breaking for Scenarios~A and~D}. The left panel gives the maximum value
              of $\tan{\beta}$ consistent with electroweak symmetry
              breaking and positive squark masses displayed as a
              function of the gravitino mass. The plot is shown with
              $b_{+}=0.08$ but the values are extremely insensitive to the
              choice of this parameter. The right panel shows three pairs of
              curves for $b_{+}=0.084, 0.087, 0.090$. For
              values of $\tan{\beta}$ between the curves the heavy
              scalar contribution at two loops to the running of $m^{2}_{U_3}$
              drives its value negative.}}
        \label{fig:lowtanbeta}
%    \end{center}
\end{figure}

The first condition to be imposed on the scenarios considered here is
correct electroweak symmetry breaking, defined
by~(\ref{eq:radmuterm}), with no additional {\mbox scalar} masses
negative. This criterion alone rules out Scenario C, with all three
generations coupling universally to the GS-counterterm and having
large {\mbox scalar}
masses. For the opposite case of no coupling to the GS-counterterm
(Scenario A) the allowed region is
displayed in Figure~\ref{fig:lowtanbeta}. In this scenario electroweak 
symmetry breaking requires $1.65 < \tan{\beta} < 4.5$, the lower bound
being the value for which the top quark Yukawa coupling develops a
Landau pole below the condensation scale. This restricted region of
the $\tan{\beta}$ parameter space is a result of the large hierarchy
between gaugino masses and scalar masses in these models and has been
observed in more general studies of the MSSM parameter space
\cite{Kolda}.

Scenario B with its split generations can exist only for 
$0.08 \leq b_{+} \leq 0.09$, where the hierarchy between the generations is
small enough to prevent the two-loop effects of the heavy generations
from driving the right-handed top squark to negative mass-squared values.
Furthermore, proper electroweak symmetry breaking in this model
requires the value of $\tan{\beta}$ to be in the uncomfortably narrow range $1.65 \leq
\tan{\beta} \leq 1.75$, making this pattern of Green-Schwarz couplings 
phenomenologically unattractive.

As for Scenario D, the large third generation masses give an
additional downward pressure on the Higgs mass-squareds in the running 
of the RGEs, allowing for a much wider allowed range of
$\tan{\beta}$. In fact, electroweak symmetry is radiatively broken in
the entire range of parameter space. However, as the value of $b_{+}$
is raised past the critical range $b_{+} \simeq 0.08$, the scalar
mass boundary values at the condensation scale start to become light
enough that the right-handed stop is again driven to negative
mass-squared values. This is shown in Figure~\ref{fig:lowtanbeta} where
the region between the upper and lower curves is excluded. While this
region expands rapidly as the beta-function coefficient is increased,
the values of the beta-function coefficient consistent with
$\alpha_{\rm str} \geq 0.04$ are nearly saturated when this effect arises.

\begin{figure}[t]
%    \begin{center}                           
\centerline{
       \psfig{file=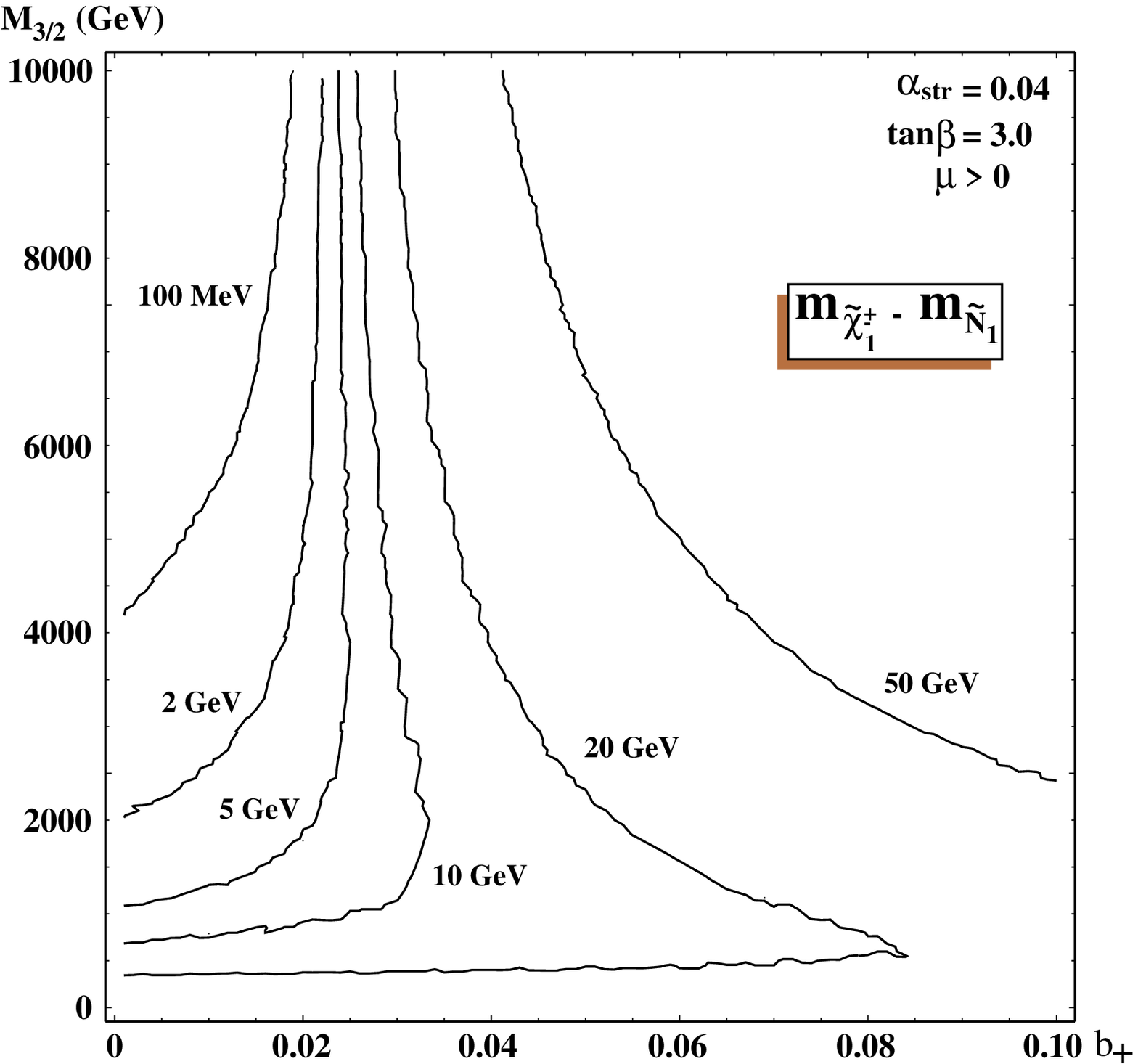,width=0.5\textwidth}
       \psfig{file=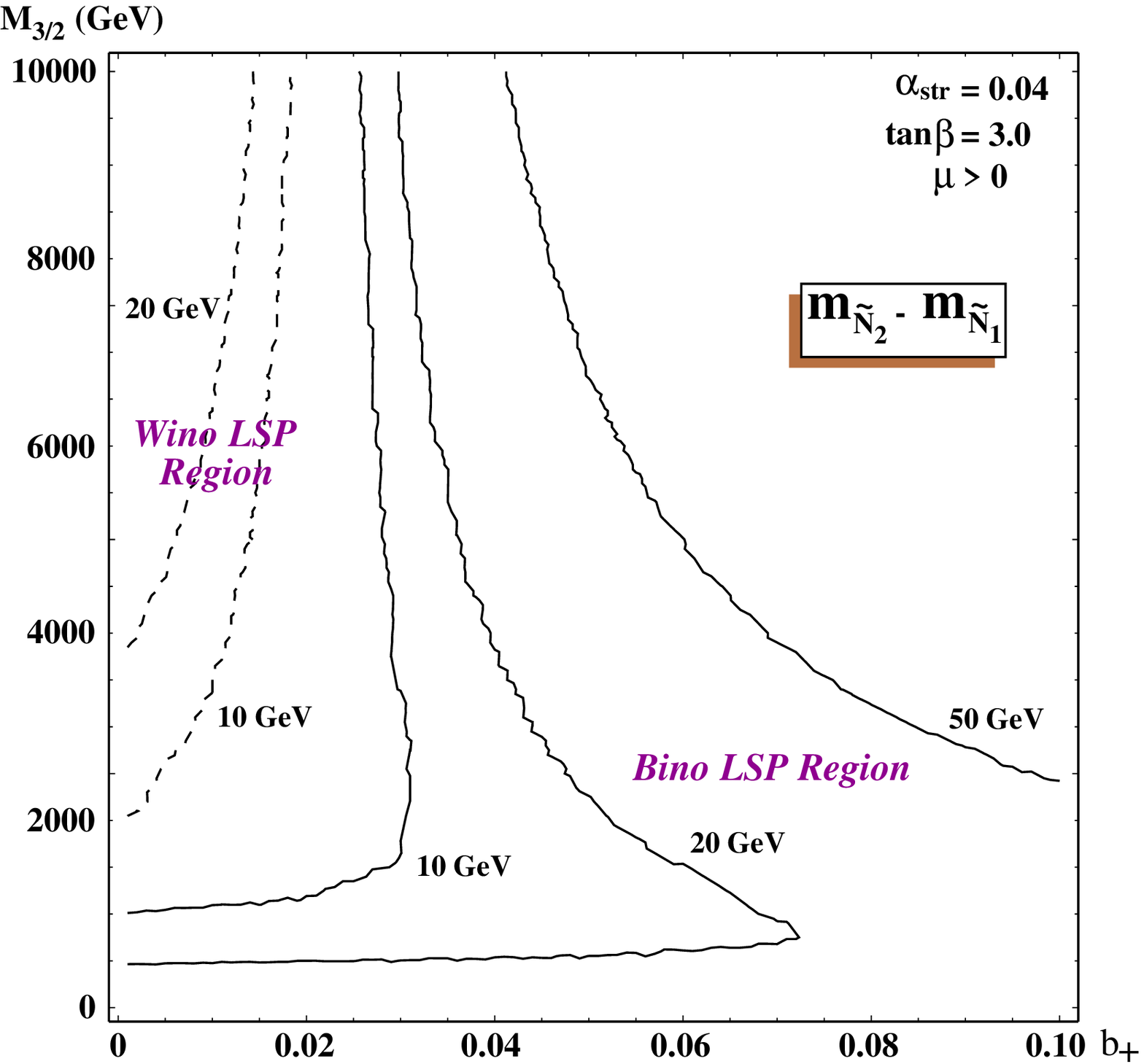,width=0.5\textwidth}}
          \caption{{\footnotesize {\bf The Physical Gaugino
                Sector in Scenario~A}. The left panel gives the mass difference
              between the $\chi^{\pm}_{1}$ and the $N_1$ in
              GeV. Typical search algorithms at colliders assume a
              mass difference at least as large as 2 GeV. The right
              panel gives the difference in mass between the two
              lightest neutralinos $N_{2}$ and $N_{1}$. Note that a
              level crossing occurs and there exists a region in which 
              the $W_{0}$ becomes the LSP, as is to be expected when
              the anomaly contribution to gaugino masses dominates.}}
        \label{fig:gauginosec}
%    \end{center}
\end{figure}

The direct experimental constraints are most binding for the
gaugino sector as they are by far the lightest superpartners in this
class of models. Typical bounds reported from collider experiments are 
derived in the context of universal gaugino masses with a relatively
large mass difference between the lightest chargino and the
lightest neutralino. For most choices of parameters in the models
studied here this is a valid assumption, but when the condensing group 
beta-function coefficient $b_{+}$ becomes relatively small
(i.e. similar in size to the MSSM hypercharge value of $b_{U(1)}=0.028$)
the pieces of the gaugino mass arising from the superconformal
anomaly~(\ref{eq:gaugino}) can become equal in magnitude to the universal
term. Here there is a level crossing in the neutral gaugino
sector. The lightest supersymmetric particle (LSP)
becomes predominately wino-like and the mass difference between the
lightest chargino and lightest neutralino becomes negligible. This
effect is
displayed in Figure~\ref{fig:gauginosec}. The experimental constraints 
as normally quoted from LEP and the Tevatron cannot be applied in the
region where the mass difference between the lightest neutralino and
chargino falls below about 2 GeV. The phenomenology of such a gaugino
sector has been studied 
recently in \cite{Gerghetta}. Note that when any scalar fields couple
to the GS-counterterm (as in Scenario~D) there is a large additional,
universal contribution to the gaugino masses at the condensation scale 
in~(\ref{eq:gaugino}). This eliminates any region with a non-standard
gaugino sector in these cases.

\begin{figure}[t]
%    \begin{center}
\centerline{
       \psfig{file=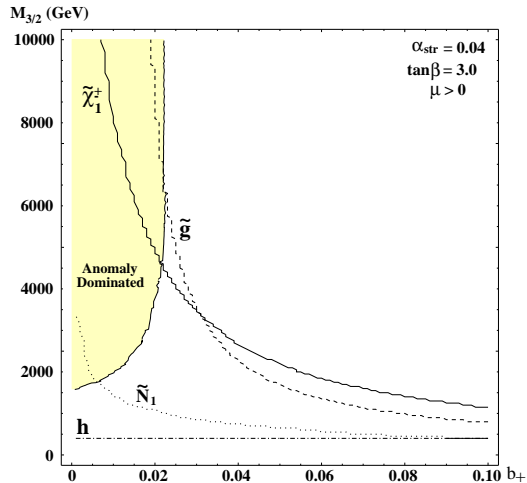,width=0.5\textwidth}}
          \caption{{\footnotesize {\bf Constraints from 
Table~\ref{tbl:massbounds} for Scenario~A}.
              Exclusion curves for lightest chargino (solid), gluino
              (dashed), lightest neutralino (dotted) and lightest Higgs 
              mass (dashed-dotted) for weak coupling at the string
              scale. The region below the curves fails to meet the
              corresponding constraint from
              Table~\ref{tbl:massbounds}. The upper left corner
              represents the region where the difference in mass
              between the $\chi^{\pm}_{1}$ and the $N_1$ falls below 2 GeV 
              and is thus not subject to the same observational
              constraints as standard minimal supergravity models.}}
        \label{fig:physmassplot}
%    \end{center}
\end{figure}

Figure~\ref{fig:physmassplot} gives the
binding constraints from Table~\ref{tbl:massbounds} for Scenario A
with $\tan{\beta}=3$ and positive $\mu$ (the most restrictive
case). The most critical constraints are for the lightest chargino and 
gluino.\footnote{The gluino mass determination takes
  into account the difference between the running mass ($M_3$) and the 
  physical gluino mass~\cite{gluino}. This difference is neglected for
  the other mass parameters.}  The effect of varying $\tan{\beta}$ on these
bounds is negligible over the range $1.65 < \tan{\beta} < 4.5$,
as its effect is solely in the variation in the Yukawa couplings appearing
at two loops in the gaugino mass evolution. The region for which the
anomaly-induced contributions to the gaugino masses make the normal
experimental constraints inoperative is represented by the shaded
region in the upper left of the figure. In general, the light
gaugino masses at the condensation scale require a large gravitino
mass (and hence, a large set of soft scalar masses since
$m_{A}=M_{3/2}$ in this scenario) in order to evade
the observational bounds coming from LEP and the Tevatron. While
current theoretical prejudice would disfavor such large soft scalar
masses, this pattern of soft parameters may not necessarily be a sign
of excessive fine-tuning \cite{heavyscalars}. Nevertheless, we refrain from 
making any statements about the ``naturalness'' of this class of
models as we have not specified any mechanism for generating the
mu-term. 

\begin{figure}[t]
%    \begin{center}
\centerline{
       \psfig{file=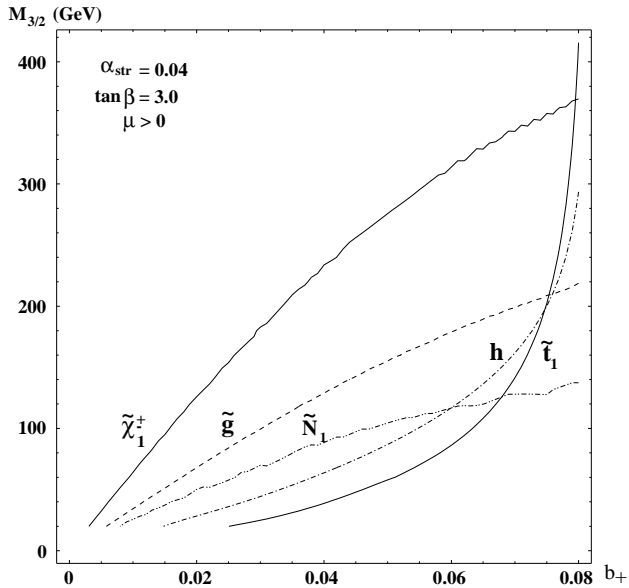,width=0.6\textwidth}}
          \caption{{\footnotesize {\bf Constraints from 
Table~\ref{tbl:massbounds} for Scenario D}.
              Exclusion curves for lightest chargino (thick solid), gluino
              (dashed), lightest neutralino (dotted), lightest Higgs 
              (dashed-dotted) and lightest stop (thin solid)
              mass. Curves are for weak coupling at the string
              scale. The region below the curves fails to meet the
              corresponding constraint from
              Table~\ref{tbl:massbounds}.}}
        \label{fig:physmassplotD}
%    \end{center}
\end{figure}

Figure~\ref{fig:physmassplotD} gives the
binding constraints from Table~\ref{tbl:massbounds} for Scenario D
with $\tan{\beta}=3$ and positive $\mu$. Note the change of scale in
both axes for these plots relative to those of Scenario A. As in
Figure~\ref{fig:physmassplot}, varying $\tan{\beta}$ over the range
$1.65 < \tan{\beta} < 40$ has a negligible effect on
the gaugino constraint contours and only a very small effect on the
contours of constant stop mass.
Here the gaugino masses start at much larger values so a lower
supersymmetry-breaking scale is sufficient to evade the bounds from LEP and the
Tevatron. Though the gravitino mass can now be much smaller, recall
that the scalars in this scenario have masses at the condensation
scale roughly an order of magnitude larger than the gravitino. Thus
the typical size of scalar masses at the electroweak scale continues
to be about 1 TeV for the first two generations and a few hundred
GeV for the third generation scalars. As
opposed to the case where all the matter fields of the observable
sector decouple from the GS-counterterm, here smaller values of the
condensing group beta-function coefficient {\em enhance} the gaugino masses
via the last term in~(\ref{eq:gaugino}).

\begin{figure}[t]
%    \begin{center}
\centerline{
       \psfig{file=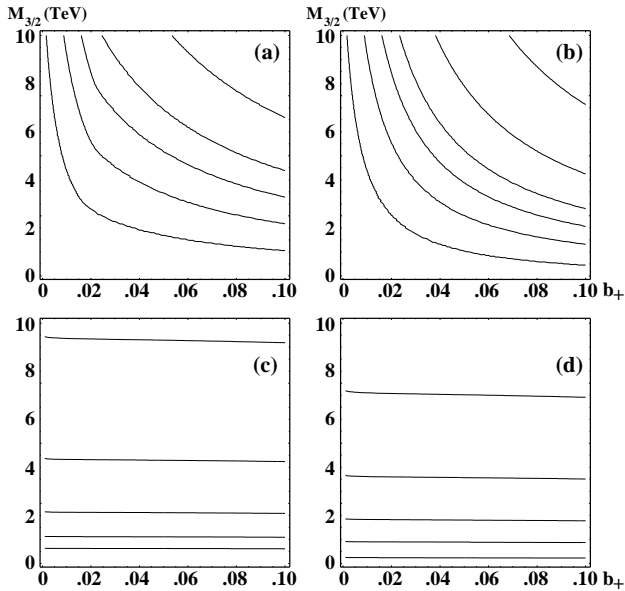,width=0.6\textwidth}}
          \caption{{\footnotesize {\bf Mass Contours for
                Scenario~A}. {\em Panel A:} Contours for the lightest
              neutralino mass of 40, 80, 120, 160 and 240 GeV. {\em
                Panel B:} Contours of lightest chargino mass of 40, 80,
              120, 160, 240 and 400 GeV. {\em Panel C:} Contours of
              lightest Higgs mass of 90, 100, 110, 120 and 130
              GeV. {\em Panel D:} Contours of lightest stop mass of
              200, 500, 1000, 2000 and 4000 GeV. All contours increase 
              from the bottom to the top of each panel.
             }}
        \label{fig:physmassplotA2}
%    \end{center}
\end{figure}

\begin{figure}[t]
%    \begin{center}
\centerline{
       \psfig{file=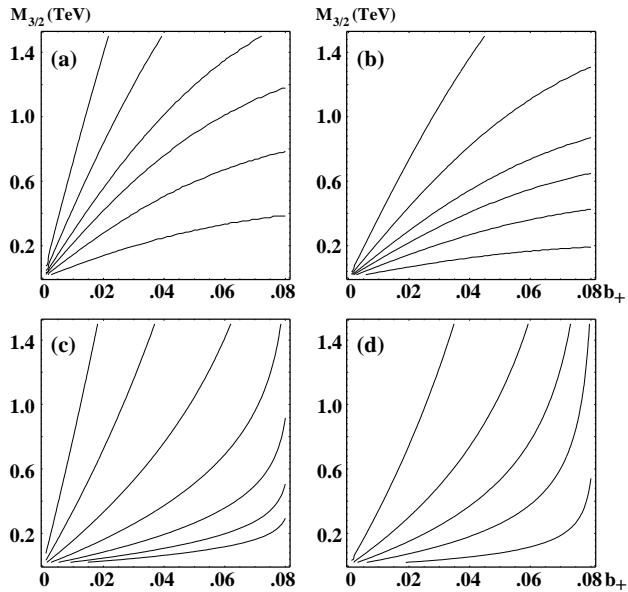,width=0.6\textwidth}}
          \caption{{\footnotesize {\bf Mass Contours for
                Scenario~D}. {\em Panel A:} Contours for the lightest
              neutralino mass of 40, 80, 120, 160, 240 and 400 GeV. {\em
                Panel B:} Contours of lightest chargino mass of 40, 80,
              120, 160, 240 and 400 GeV. {\em Panel C:} Contours of
              lightest Higgs mass of 80, 90, 100, 110, 120, 130 and 140
              GeV. {\em Panel D:} Contours of lightest stop mass of
              200, 500, 1000, 2000 and 4000 GeV. All contours increase 
              from the bottom to the top of each panel.
             }}
        \label{fig:physmassplotD2}
%    \end{center}
\end{figure}

We end this section by giving mass contours for the lightest Higgs,
chargino, neutralino and
top-squark for $\tan{\beta}=3$ and positive $\mu$ for Scenarios~A
and~D in Figures~\ref{fig:physmassplotA2} and~\ref{fig:physmassplotD2}, 
respectively.

\section{Gauge Coupling Unification}
\label{sec:gaugeunify}
In Section~\ref{sec:implication} the possibility of
larger values of the unified coupling constant $g^{2}_{\rm str}$ at
the string scale was considered in a very general way. It is well
known~\cite{unification} that the apparent unification of coupling constants at a scale
$\Lambda_{\rm MSSM} \approx 2 \times 10^{16}$~GeV, assuming only the
MSSM field content, is at odds with the string prediction that
unification must occur at a scale given by 
\begin{equation}
M_{\rm string}^{2} = \lambda g_{\rm string}^{2}  M_{\rm Planck}^{2}
\label{eq:stringscale}
\end{equation}
where $\lambda$ represents the (scheme-dependent) one-loop correction from heavy string
modes. In \cite{ModInv} this factor was computed for the $\overline{MS}$
scheme and it is given by
\begin{equation}
\lambda = \frac{1}{2}\(f+1\)e^{g-1}.
\label{eq:factor}
\end{equation}
For the vacua considered in this work this parameter is typically
$\lambda \sim 0.19$. 

Even after taking into account one-loop string corrections there is
still an order of magnitude discrepancy between the scale of unification
predicted by string theory and the apparent scale of unification as
extrapolated from low energy measurements under the MSSM framework. One possible 
solution to the problem is the inclusion of additional matter
fields in incomplete multiplets of SU(5) at some intermediate scale
which will alter the running of
the coupling constants, causing them to converge at some value higher
than $\Lambda_{\rm MSSM}$~\cite{newmatter}. These solutions tend to
involve slightly larger values of the coupling constant at the string scale
than that of the MSSM ($\alpha_{\rm MSSM}^{-1} \approx 24.7$).

In the model in question here, the intermediate scale ($\Lambda_{\rm
  cond}$) at which this
additional matter might appear is not independent of the
scale of the superpartner spectrum ($\Lambda_{\rm SUSY} \sim
M_{3/2}$), but the two are in fact related by
equation~(\ref{eq:gravmass}). Thus if we assume this additional matter 
has a typical mass of the condensation scale, each point in the  $\lbr b_{+},
M_{3/2} \rbr$ plane can be tested for potential compatibility with
string unification given a certain set of additional matter fields. We 
will not specify the origin of these fields (though such incomplete
multiplets are not uncommon in string theory compactifications), but
merely posit their existence with masses on
the order of the condensation scale.

Our procedure for carrying out this investigation is similar to that
used in the literature by a number of authors \cite{gaugeRGE}. The standard
model coupling constants $\alpha_{3}$, $\alpha_{2}$ and $\alpha_{1}$
are determined from $\alpha_{\rm EM}\(M_Z\)=1/127.9$, $\alpha_{3}\(M_Z\)=0.119$ and
$\sin^{2}\theta_{\rm EW}\(M_Z\)=.23124$ and these $\overline{MS}$
values are converted to the $\overline{DR}$ scheme. As we
will not be concerned with performing a precision survey, these
coupling constants are run at one loop from their values at
the electroweak scale using only the standard model field content up
to the scale $\Lambda = M_{3/2}$. At this scale the entire supersymmetric
spectrum is added to the equations until the scale $\Lambda = \Lambda_{\rm cond}$ is
reached. Here incomplete multiplets of SU(5) are added and the
couplings are run to the scale at which the SU(2) and U(1) fine
structure constants coincide. This scale will be defined as the string scale.

We now require $\alpha_{3} =\alpha_{2} =\alpha_{1}$ at this
scale and invert equation~(\ref{eq:stringscale}) to find the implied
Planck scale. Consistency requires that this value be the reduced
Planck mass of $2.4 \times 10^{18}$ GeV {\em and} that the QCD gauge
coupling, when the renormalization group equations are solved in the
reverse direction, give a value for $\alpha_{3}$ at $\Lambda = M_{Z}$ within
two standard deviations of the measured value.\footnote{It is worth
  remarking that even the celebrated supersymmetric SU(5) unification of
  couplings fails to predict the strong coupling at the electroweak
  scale at the level of two sigma and calls for a rather large value
  of $\alpha_{3}\(M_{Z}\)$\cite{gaugeRGE}. This is usually taken as an 
  indication of the size of
  model-dependent threshold corrections. We
  therfore demand no more from the models considered here.} 

\begin{figure}[t]
%    \begin{center}
\centerline{
       \psfig{file=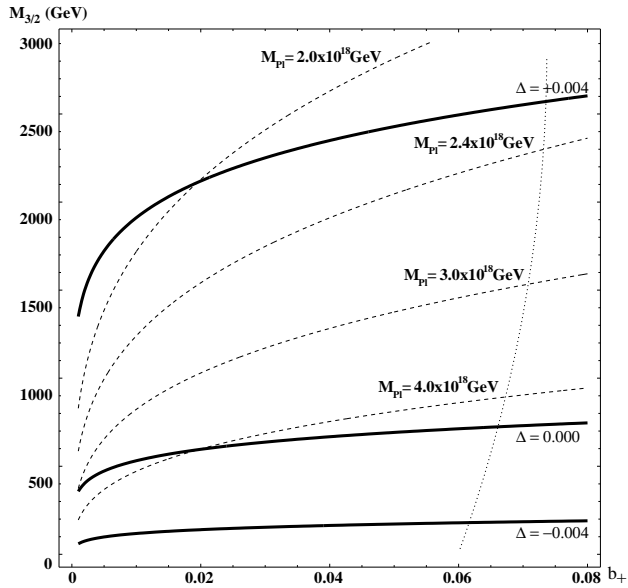,width=0.6\textwidth}}
          \caption{{\footnotesize {\bf Gauge Coupling Unification}.
           Results of adding one pair of $\(Q,\bar{Q}\)$ and two pairs 
           of $\(D, \bar{D}\)$ at the condensation scale. Contours of
           constant implied Planck mass are overlaid on
           the region for which $\Delta=\alpha^{\rm
             RGE}_{3}-\alpha^{\rm obs}_{3}$ is within the two-sigma
           experimental limit of $\delta\alpha=\pm 0.004$. The
           dotted line represents the maximum value of $b_{+}$
           consistent with $M_{3/2}\leq10$ TeV and the RGE determined
           string coupling. The values of $\alpha_{str}$ here range
           from $0.044$ at the $\Delta=+0.004$ contour to $0.050$ at the
           $\Delta=-0.004$ contour.}}
        \label{fig:unify1}
%    \end{center}
\end{figure}

The results of the analysis for a typical choice of extra matter
fields are shown in Figure~\ref{fig:unify1}, where
a pair of vector like $\(Q,\bar{Q}\)$ and two pairs of vector-like
$\(D,\bar{D}\)$'s are introduced at the condensation scale with quantum
numbers identical to their MSSM counterparts. The two sigma window
about the current best-fit value of $\alpha_{3}$ can indeed accomodate
a consistent Planck mass while allowing for perturbative unification
of gauge couplings. From this base configuration additional {\bf 5}s
and {\bf 10}s of SU(5) can be added at will to increase the value of
the unified coupling at the string scale, but the contours of constant 
implied Planck mass shown in Figure~\ref{fig:unify1} will not move significantly.
While these combinations of
matter fields have been known to allow for gauge coupling unification
for some time \cite{newmatter}, the relationships~(\ref{eq:gravmass})
and~(\ref{eq:stringscale}) between the various scales involved makes
this a nontrivial accomplishment for this class of models.

\section*{Conclusion}
The preceeding pages should be cause for guarded optimism with regard
to string phenomenology. The initial challenge of dilaton
stabilization has been met without resorting to 
strong coupling in the effective field theory nor requiring delicate
cancellations. Reasonable values of the supersymmetry-breaking scale
can be achieved over a fairly large region of the parameter space, but 
a given combination of coupling strength at the string scale and
hidden sector matter content will single out a tantalizingly small
slice of this space. These successful combinations do not destroy the
potential solutions to the coupling constant unification problem by
the introduction of additional matter at the condensation
scale. Tighter restriction on the hidden sector will require more
precise knowledge of the size of Yukawa couplings in the corresponding 
superpotential.

Requiring a vacuum
configuration which gives rise to successful electroweak symmetry
breaking seems to demand that either the Green-Schwarz counterterm be
independent of the matter fields or that all matter fields couple in a universal 
way but that the Higgs fields are distinct. The pattern of soft
supersymmetry-breaking parameters in the former case
pushes the theory towards large gravitino masses and very low
values of $\tan{\beta}$. The low gaugino masses relative to scalar
masses favors larger beta-function coefficients for the condensing
group of the hidden sector, while smaller values may result in
phenomenology in the gaugino sector similar to that of the ``anomaly
dominated'' scenarios.

In the latter case a proper vacuum configuration and weak coupling at
the string scale leave the value of $\tan{\beta}$ free to take its
entire range of possible values. Larger beta-function coefficients for 
the condensing group allow a promising region with relatively light
scalar partners of the third-generation matter fields and light gauginos.

A more realistic
model may alter these results to some degree and uncertainty remains in 
the general size and nature of the Yukawa couplings of the hidden
sector of these theories. Nevertheless this survey suggests that
eventual measurement of the size and pattern of supersymmetry breaking 
in our observable world may well point towards a very limited choice
of hidden sector configurations (and 
hence string theory compactifications) compatible with low energy phenomena.

\vskip .5cm
\section*{Acknowledgements}
\vskip .5cm

We than Pierre Bin\`etruy, Hitoshi Murayama and Marjorie Shapiro for
discussions. This work was supported in part by the Director, Office of 
Energy Research, Office of High Energy and Nuclear Physics, Division 
of High Energy Physics of the U.S. Department of Energy under 
Contract DE-AC03-76SF00098  and in part by the National Science 
Foundation under grant PHY-95-14797 and PHY-94-04057.
\hspace{0.8cm}

\pagebreak

\end{document}